\documentstyle[psfig,aps,preprint]{revtex}

\def\m@thcombine#1#2{%
  \setbox0=\hbox{$#1$}
  \setbox1=\hbox{$#2$}
  \ifdim\wd0>\wd1
    \setbox0=\hbox to\wd1{\hss\box0\hss}
  \else
    \setbox1=\hbox to\wd0{\hss\box1\hss}
  \fi
  \mathop{\vcenter{
    \offinterlineskip\box0\box1}}}
\def\lesim{\m@thcombine<\sim}
\def\gesim{\m@thcombine>\sim}

\newcommand{\ket}[1]{| {#1} \rangle}
\newcommand{\bra}[1]{\langle {#1} |}

\newcommand{\eps}{\epsilon}
\newcommand{\calJ}{{\cal{J}}}
\newcommand{\del}{\partial}

\begin{document}

\title{ Violation and persistence of the K-quantum number \break
in warm rotating nuclei }

\author {M. Matsuo$^{1}$, T. D{\o}ssing$^{2}$, 
A. Bracco$^{3}$,  G.B. Hagemann$^{2}$, \\
B. Herskind$^{2}$, S. Leoni$^{3}$, E. Vigezzi$^{3}$}

\address{$^1$ Graduate School of Science and Technology, Niigata
University, Niigata 950-2181, Japan} \address{$^2$ Niels Bohr
Institute, University of Copenhagen, DK2100 Copenhagen \O, Denmark}
\address{$^3$ INFN sez. Milano, and Department of Physics, University
of Milano,Milan 20133, Italy}

\maketitle
\vspace{-5mm}
\begin{abstract} 
\setlength{\baselineskip}{6.5mm}
The validity of the K-quantum number in rapidly rotating 
warm nuclei is investigated as a function of thermal excitation energy
$U$ and angular momentum $I$, for the rare-earth nucleus $^{163}$Er.
The quantal eigenstates   
are described with a shell model which combines a 
cranked Nilsson mean-field and a residual two-body interaction, 
together with a term which   takes into account 
the angular momentum carried by the K-quantum number 
in an approximate way. 
K-mixing is produced by the interplay of the Coriolis interaction and 
the residual interaction; it is  weak in the 
region of the discrete rotational bands ($U \lesim 1$MeV), but it gradually 
increases until the limit of complete violation 
of the K-quantum number is approached around $U\sim 2 - 2.5$ MeV.
The calculated matrix elements between bands having different K-quantum
numbers decrease exponentially
as a function of $\Delta K$, in qualitative agreement with recent data.

\vspace{5mm}


\end{abstract}

\vspace{10mm}
\begin{minipage}[h]{15cm}
\begin{small}
\noindent
Corresponding author: 

\vspace{-4.5mm}
\noindent
Masayuki Matsuo

\vspace{-4.5mm}
\noindent
Graduate School of Science and Technology, Niigata University

\vspace{-4.5mm}
\noindent
Ikarashi Ninocho, Niigata 950-2181, Japan

\vspace{-4.5mm}
\noindent
e-mail: matsuo@nt.sc.niigata-u.ac.jp

\vspace{-4.5mm}
\noindent
telefax: +81-25-263-3961

\end{small}
\end{minipage}

\vfill
\break

\section{Introduction}

Recent experimental studies of the $\gamma-$decay following
fusion reactions have been able to explore in great detail the
$\gamma$-ray spectra of well deformed nuclei. For example, in the
case of the axially symmetric nucleus $^{163}$Er,
about 15 rotational bands have
been resolved up to rotational frequency around 500 keV \cite{er163},
and it has been confirmed that the K-quantum number \cite{BM2}
 is a basic  quantity
in classifying the rotational bands.
These bands cover a region of rather low
energy above the yrast line, and
it has been possible to classify interactions at band crossings
in relation to the differences between
the K-quantum number of the crossing bands \cite{Dossing99}. It was
found that significant K-selection rules still exist around
angular momentum $I \sim 25$, although much weaker than those
valid at low $I$ close to the band heads.

With the same experiments, one can also analyze the quasicontinuum 
spectra of $\gamma$-rays
emitted from states of somewhat higher thermal excitation energy
\cite{Dossing99,Herskind92,Dossing96,Dossing01}.
In this case, one can probe the feeding
into specific configurations by gating on selected low lying transitions.
In particular, the study of the covariance of spectra gated by
high-K and low-K bands has indicated a certain persistence of
K-selection rules \cite{Bosetti96}.
As it has been clearly stated by B. Mottelson \cite{Mottelson}, the
question
of K-quantum number violation in thermally excited states
is a key issue in the study of
the transition between ordered and chaotic many-nucleon motion
caused by the residual interaction and high level density
\cite{Chaos,weid,zelev}. 
The information gathered by fusion experiments  supplements in an
important way
that  available through the study of the $\gamma-$decay from neutron
resonances.

Generally, the complex nature of states at neutron resonances may amplify 
symmetry breaking, parity violation being the most striking
example \cite{Chaos}.
Theoretical estimates based on the Gaussian Orthogonal Ensemble
model\cite{RMT} and the Coriolis force definitely predict a complete
K-violation at the neutron resonance states in the case of neutron
resonances\cite{Barrett,Mottelson2}.
The model applied in this paper is in line with
this prediction, since we find that our results can be reasonably well
described by the GOE, for  thermal excitation energy $U$ above about
2 MeV, both with regard to rotational transition matrix
elements and level spacings\cite{Matsuo93,AAberg,Matsuo97b}.
On this basis, it is indeed surprising that an experimental
analysis\cite{ngamma} of transitions
from neutron resonance states finds significant K-hindrance effects
in the $\gamma$-decay following thermal neutron capture. This question
of the K-quantum number of neutron resonance states has been discussed
from various theoretical viewpoints\cite{Barrett,hansen,soloviev}.

In the present paper, we give a theoretical description of the
K-quantum number in rotating nuclei, focusing on the evolution
with increasing angular momentum and heat energy, presenting
results for the nucleus $^{163}$Er.

We formulate our description on the basis of the
cranked-shell model that has been previously developed to
describe the rotating compound states and the associated rotational
damping phenomena\cite{Matsuo97}, including a two-body residual interaction.  
However, we shall introduce a modification  to the standard cranking
hamiltonian, that allows a realistic description of high-K states.
The cranking, or Coriolis, term provides a weak breaking of the
K-quantum number. This weak breaking is in turn amplified by the
mixing between bands, caused by the residual interaction. It is
to be noted that the residual interaction we have adopted would not
by itself induce any K-mixing.

\section{Cranked shell model and high-K states}

\subsection{Cranked mean field and residual interaction}

Assuming that an axially deformed nucleus rotates about a principal axis 
(usually taken as the $x$-axis) with a constant rotational frequency,
the standard cranking model\cite{Bengtsson79} 
can describe the rotational bands associated to a given
intrinsic configuration. 
We start with 
the cranked Nilsson single-particle 
Hamiltonian $h_{cr}=h_{Nilsson} - \omega_xJ_x$. 
The rotational frequency $\omega_x$ is 
chosen so that 
the mean expectation value of the angular momentum $J_x$ along the
rotational $x$-axis is equal to the spin $I$. Many-particle
many-hole ($n$p-$n$h) configurations in the cranked Nilsson Hamiltonian
are produced with use of the common rotational frequency $\omega_x(I)$,
and adopted as an unperturbed shell model basis for the spin $I$,
which we denote by $\ket{\mu(I)}$. The basis states are regarded as 
unperturbed rotational bands. Since the angular momentum variable in the
cranking model is the component $I_x$ along the rotational $x$-axis, 
we can construct the laboratory energy $E_\mu(I)=E_\mu(I_x)$ of 
the basis state $\mu$
by identifying 
the angular momentum along the rotational $x$-axis
$I_x=\bra{\mu}J_x \ket{\mu}$
with the total angular momentum $I$. 
The laboratory energy $E_\mu(I)$ is a sum of 
occupied cranked-Nilsson routhian energies, 
the Strutinsky correction with the rotating liquid drop energy, 
and the energy correction due the transformation
to the laboratory frame. 
In addition, we introduce a residual interaction 
among the basis configurations, 
adopting the surface-delta interaction (SDI)\cite{SDI}. 
The shell model Hamiltonian
thus constructed at spin $I$ reads
\begin{equation}
H_{\mu\mu'}(I) =E_\mu(I)\delta_{\mu\mu'}+V_{\mu\mu'}(I), \label{smHaml}
\end{equation}
where $V_{\mu\mu'}(I)$ is the matrix elements of the SDI.
The residual interaction is essential to describe the compound states at
finite thermal energy and the rotational damping since both are
caused by mixing among the basis configurations\cite{Lauritzen86}
. The states near the
yrast line, on the other hand, are less influenced by the residual
interaction, and they survive as discrete rotational bands.
The details of the cranked shell model Hamiltonian (\ref{smHaml}) 
are given in Ref.\cite{Matsuo97}.  

\subsection{Hamiltonian including the K-quantum number}

In deriving the Hamiltonian (\ref{smHaml}) we have neglected the 
K-quantum number by applying the angular momentum relation 
$I=I_x$, which should be modified 
if the K-quantum number is taken into account. 

Let us now discuss how the main effects associated to the 
K-quantum number can be incorporated. 
If we consider a rotational band $i$ carrying the K-quantum number 
$K_i$, the angular momentum about the rotational $x$-axis will be given by
$I_x=\sqrt{I^2 - K_i^2}$. 
Denoting by $E^{crank}_i(I_x)$ the rotational energy acquired by the band 
 cranking  about the $x$-axis, 
the rotational energy taking into account the K-quantum number  
is given by  $E_{i,K_i}(I)
=E^{crank}_i (I_x=\sqrt{I^2 - K_i^2})$. 
In the case of high spin
states, that is $I \gg K_i$, or  if the dependence on $I_x$ is quadratic as
expected from a rigid-body rotation, 
the above equation can be approximated as 
\begin{equation}
E_{i,K_i}(I) =E^{crank}_i (I_x=I) -K_i^2/2\calJ_i,   \label{kcorrection}
\end{equation}
where 
$
\calJ_i=I_x(\del E^{crank}_i/\del I_x)^{-1}=I_x/\omega_x   
$
is the kinematic moment of inertia of the
band.  Note that the energy correction $- K_i^2/2\calJ_i$ for small
angular momenta converges toward the corresponding term in Bohr's
collective rotational energy 
$E_{i,K_i}(I)= \{I(I+1)-K_i^2\}/2\calJ_i +E_i^{intr}$
($E_i^{intr}$ being the intrinsic excitation energy)  \cite{BM2}. 
The first term $E^{crank}_i (I_x=I)$
corresponds to our previous cranked
shell model Hamiltonian,  Eq.(\ref{smHaml}). The second term  
$-K_i^2/2\calJ_i$ represents the effect of the K-quantum number
on the rotational energy, which we want to include in the present calculations.

In accordance with these considerations, we modify the 
shell model Hamiltonian as 
\begin{equation}
H_{\mu\mu'}(I) =E_\mu(I)\delta_{\mu\mu'}+V_{\mu\mu'}(I)
-(J_z^2)_{\mu\mu'}/2\calJ_{\mu\mu'}  \label{smHamlK}
\end{equation}
The last term is an operator form of the energy correction $- K_i^2/2\calJ_i$,
where $J_z$ is
the angular momentum operator  of the 
constituent nucleons along the symmetry axis $z$,
and 
$\calJ_{\mu\mu'}=(\calJ_\mu+\calJ_{\mu'})/2$ is the kinematic moment of
inertia of the np-nh basis configurations. 
Note that $J_z^2$ has off-diagonal matrix elements
$\mu \neq \mu'$ since the unperturbed states $\ket{\mu(I)}$ do not 
necessarily have a good K-quantum number. 
The new term in the Hamiltonian favors the states with high values of the 
K-quantum number, as will be illustrated below.
Appendix A describes how high-K (or rather high-K$^2$) states
are generated from the signature basis by the Hamiltonian
(\ref{smHamlK}).

It is possible to relate our cranked shell model to the
tilted-axis cranked mean-field (TAC) model \cite{TAC}, which 
has been used
to describe the high-K rotational bands near the yrast line.
If one makes a
mean-field approximation to the $J_z^2$-term, it reads
$J_z^2/2\calJ\sim (\left<J_z\right>/\calJ)J_z=\omega_z J_z$, which
corresponds to the second part of the generalized cranking term
$\omega_x J_x + \omega_z J_z$ in the TAC model.
Note however that the self-consistency between the directions of the 
angular momentum vector and the rotational
frequency vector, required by the tilted rotation,
is achieved approximately
but not exactly in the present model 
since a common cranking rotational frequency $\omega_x$ is
ascribed in our case for all the states at a given spin. 
This approximation affects also the energy eigenstates obtained 
by diagonalizing the Hamiltonian (\ref{smHamlK}).
The TAC mode is more consistent in this respect since the self-consistency
is achieved for each rotational band.
Our treatment enables, on the other hand, a description of
configuration mixing in compound states caused by the residual two-body
interaction.

\section{Mixing of K-quantum number}

\subsection{Effects on low-lying bands }

Numerical calculations have been performed for $^{163}$Er. 
In the calculations, we employ 4000 np-nh basis states with the lowest
excitation energies to diagonalize the Hamiltonian (1) for each
$I^\pi$.  The truncation corresponds to a cut-off in excitation energy of
approximately 4 MeV.  This larger basis (compared to
Ref.\cite{Matsuo97}) is necessary to produce  the high-K 
states whose energy is lowered by the $J_z^2$-term after the basis
selection has been made. The deformation parameters
$\eps_2=0.252$, and $\eps_4=-0.004$ are taken from the tilted axis
cranking analysis of the same nucleus\cite{Brockstedt-er163}, while
the other model parameters are the same as those of
Ref.\cite{Matsuo97}.

\begin{figure}[t]
\centerline{\psfig{figure=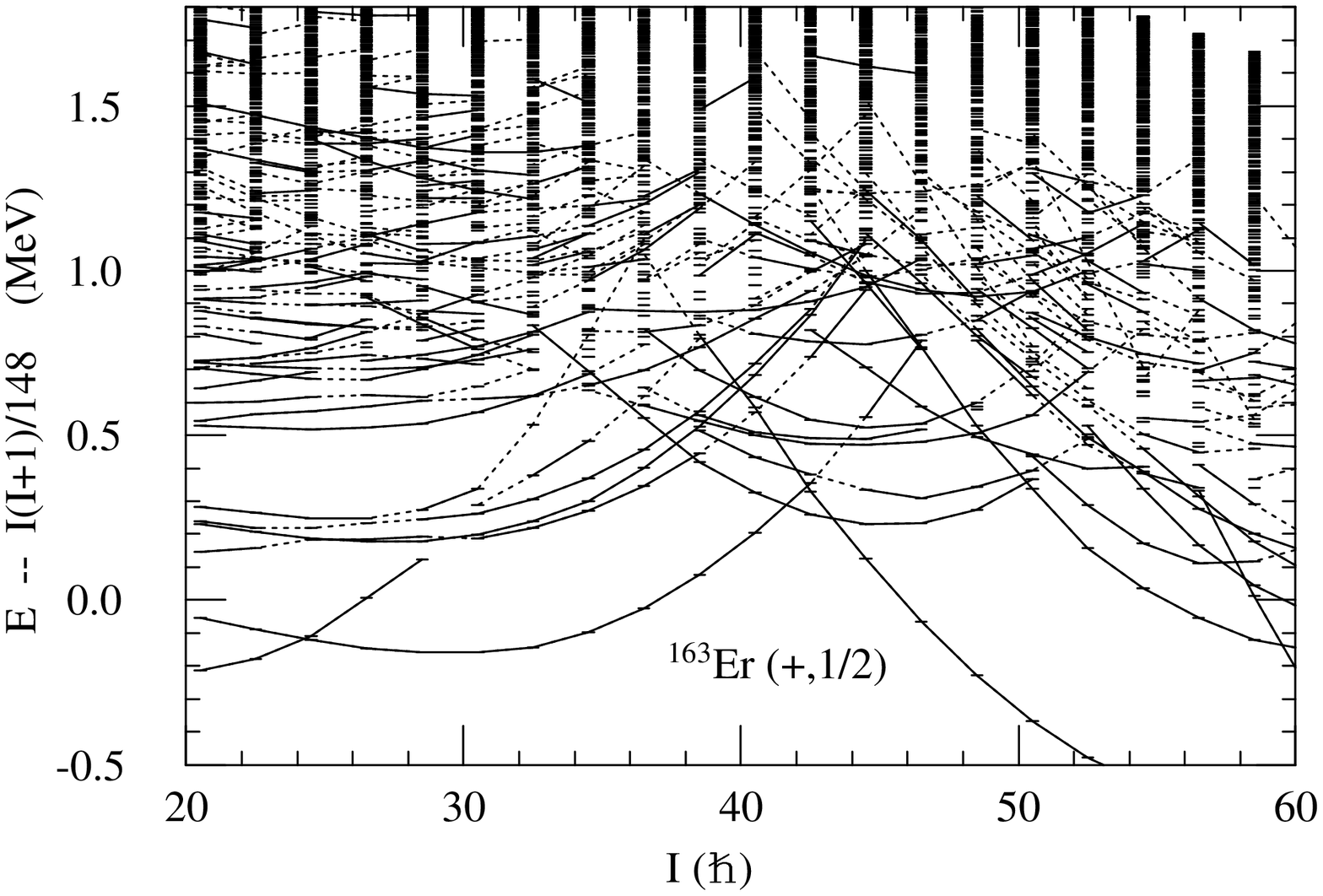,width=140mm}}
\caption{The calculated energy levels in $^{163}$Er with parity and 
signature 
$(\pi, \alpha)=(+,1/2)$. A reference rotational energy
 $I(I+1)/148$ MeV is subtracted. The solid (dotted) 
lines connecting energy levels
indicate stretched E2 transitions exhausting more than $71\%$ 
($50\%$) of 
the full rotational E2 strength.}
\label{Bands}
\end{figure}

\begin{figure}[t]
\centerline{\psfig{figure=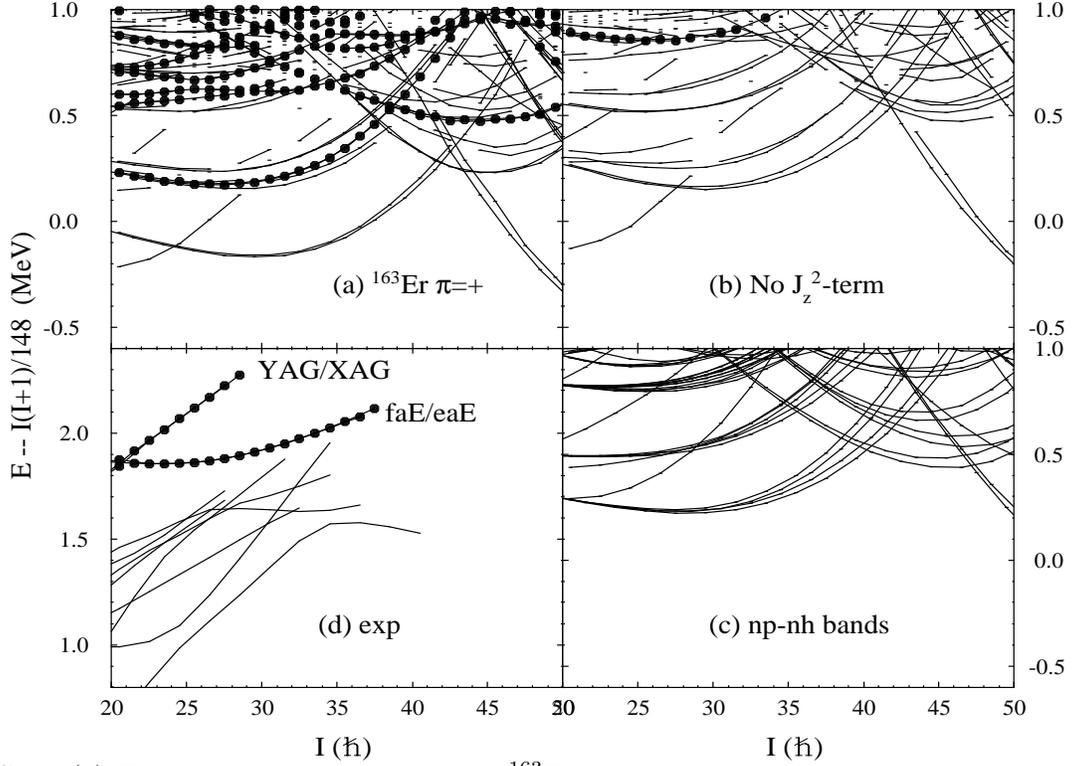,angle=-90,width=140mm}}
\caption{
(a) The calculated energy levels in $^{163}$Er with positive
parity,  with emphasis on high-K states with $K_i>8$ marked with the
filled circle. Both signatures $\alpha=\pm 1/2$ are included. The
solid line connects  the E2 transitions with more than $71\%$.
(b) The same as (a), but the $J_z^2$-term is neglected.
(c) The same as (a), but the unperturbed np-nh cranked mean-field bands
are displayed.
(d) The experimental rotational bands with positive parity[1].
The high-K bands are marked with the filled circle, and the labels
represent the configurations assigned in the experiment (see text).
} 
\label{Bands2}
\end{figure}

The energy levels calculated in the spin range  $I= 20-60$
are shown in Fig.\ref{Bands}, where 
the strong stretched E2 transitions forming rotational band 
structures are evidenced. 
Many of the states whose excitation energy $U$ relative to yrast
is smaller than about 1 MeV form rotational bands whereas the rotational
band structure gradually disappears as $U$ becomes larger  than  $\sim 1$MeV,
indicating that the rotational damping sets in around this 
excitation energy.
We can define the average value of $K$ for each energy level $i$ 
by $K_i=\sqrt{\bra{i}J_z^2\ket{i}}$
in terms of the expectation value of $J_z^2$. It is found that
many high-K states with a large value of $K_i$, e.g., $K_i>8$, are produced.
Examples for the positive parity states are shown in Fig.\ref{Bands2}(a). 
The lowest energy positive parity high-K states at excitation energy
$U\approx 0.4$MeV form a rotational band extending up to $I\sim 89/2$
with essentially zero signature splitting. 
It has a large value of  $K_i$, $K_i \approx 10$, and its eigenfunction is 
predominantly 
$(n[523]5/2)(p[404]7/2)(p[523]7/2)^{-1}$ with the combination of
$K=5/2+7/2+7/2=19/2$. 
The role of the new $J_z^2$-term in the Hamiltonian (3) can be appreciated 
comparing Fig.\ref{Bands2}(a) to  \ref{Bands2}(b), where the $J_z^2$-term
is neglected,  disregarding the contribution
of the K-quantum number to the angular momentum. As a consequence, in 
Fig. \ref{Bands2}(b)  there are essentially no low-lying high-K
states. Fig.\ref{Bands2}(c) shows the np-nh mean-field basis bands,
where both the residual interaction and the $J_z^2$-term are neglected.
Eight np-nh basis bands located at $E-I(I+1)/148 \approx 0.8$MeV 
at spin $I \sim 20-25$ exhibit an approximate octet degeneracy, arising
from the $(n[523]5/2)(p[404]7/2)(p[523]7/2)^{-1}$ configurations where
each cranked Nilsson orbit has approximate two-fold degeneracy
with different signatures. The $J_z^2$-term resolves the octet degeneracy
and produces the lowest high-K state $K_i \approx 19/2$ 
in Fig.\ref{Bands2}(a). The energy shift due to the $J_z^2$-term
amounts to several hundreds of  keV. 

The lowest high-K band in Fig.\ref{Bands2}(a) corresponds to 
the positive parity high-K rotational band (faE/eaE) 
observed in the experiments (Fig.\ref{Bands2}(d)), 
for which a similar quasiparticle configuration and $K=19/2$ have been 
assigned \cite{er163}. The calculated excitation energy of this
band relative to yrast is about 350-500 keV, which is
in fair agreement with the experiment concerning the highest spin
region $I\approx 30-36$. 
The second positive parity high-K band (YAG/XAG) observed 
in the experiment has an excitation energy $U=1.1-1.3$MeV.
In the calculation, a second  excited
high-K bands with positive parity lies around $U=0.8-0.9$MeV, showing that  
we here have an overall description of the high-K bands near yrast.
However, specific properties of individual 
rotational bands, including both high-K and low-K bands,
are not well reproduced. The calculated second high-K band has
slightly different configurations from that of the observed one while
both involve the deformation-aligned neutron orbit $(n[505]11/2)$. 
Also,
alignments of i$_{13/2}$ quasi-neutrons and h$_{11/2}$ quasi-protons observed
in the experiments are missing. 
These disagreements may be traced back mostly to the insufficient treatment 
of  pairing correlations, which are only partly  taken into account 
through the shell model diagonalization in the present model.
A pairing gap of 0.8-0.9MeV is adopted both for neutrons and protons
in the tilted axis cranking (TAC) calculation\cite{Brockstedt-er163}, 
which provides a better description of near-yrast rotational bands. 
Note also that the TAC treats the self-consistency in the rotational 
angular momentum vector more precisely.

It is noted that deformation aligned 
single-particle orbits having
sizable value of $\Omega_i$ (the approximate cranked Nilsson 
quantum number of $J_z$) lying near the Fermi surface 
play important roles to produce high-K states. 
Their contribution depends on the neutron and proton numbers, and
deformations. Concerning the neutrons, the orbits 
$n[523]5/2$ and $n[505]11/2$ are relevant. 
If we consider slightly heavier nuclei around 
$^{168}$Yb,  contributions from these orbits are less effective, and
the same cranked shell model produces a smaller number of high-K states.

\subsection{K-mixing as a function of thermal excitation energy and spin}

\begin{figure}[t]
\centerline{\psfig{figure=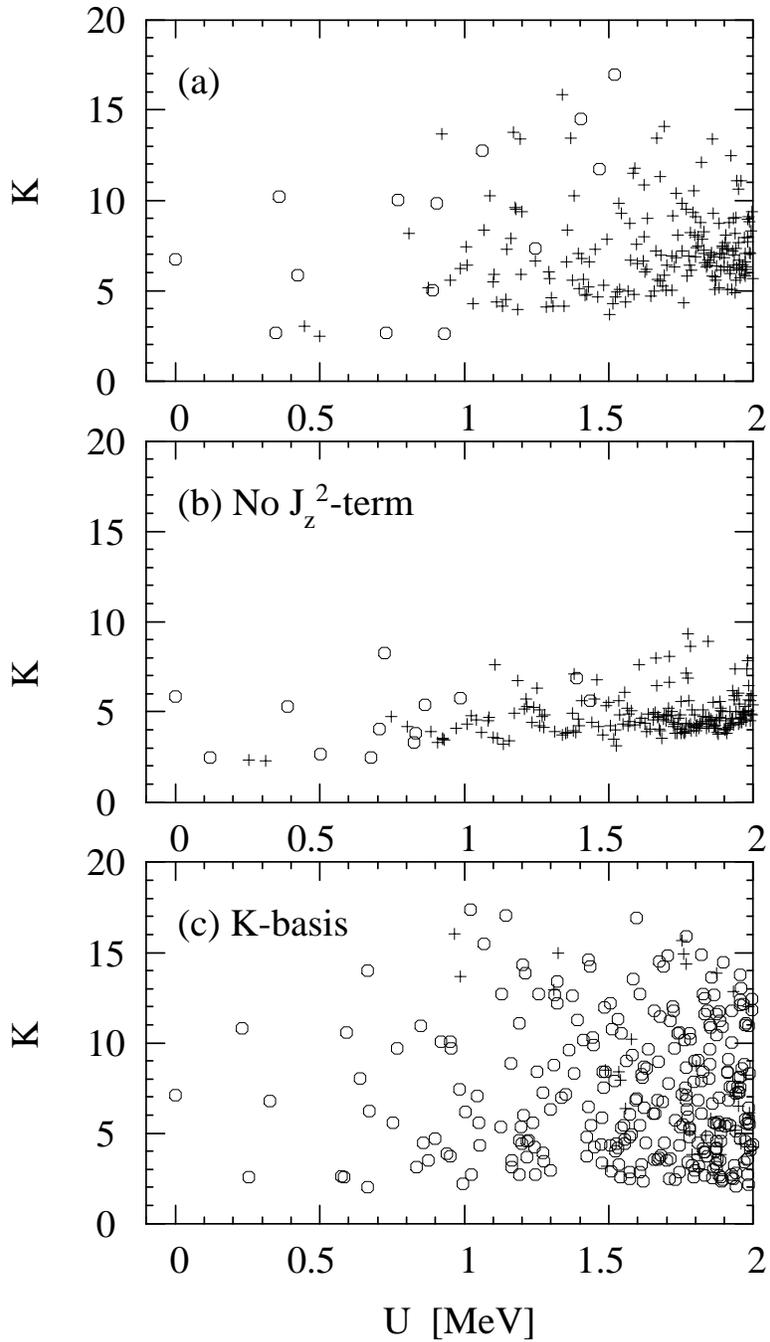,width=100mm}}
\caption{The K values of the calculated states with $I^\pi={61 \over 2}^{+}$.
The states forming rotational bands are marked with open circle,
whereas the states with rotational damping are shown by crosses. 
(a) The full shell model Hamiltonian
including the $J_z^2$-term is diagonalized. (b) The $J_z^2$-term is neglected.
(c) The $J_z^2$-term is kept but the residual two-body interaction is 
neglected to produce the K-basis states.}
\label{Kvalue}
\end{figure}

Fig.\ref{Kvalue}(a) plots the value of $K_i$ for the 
energy levels up to $U=2$ MeV at $I^\pi=61/2^+$. The energy
levels are marked with different symbols depending on whether
they are connected with strong E2 transitions to spin $I-2$ or $I+2$ 
(cf. Fig.\ref{Bands}).
There exist several high-K rotational bands with $K \approx 10-15$
below $U\sim 1.5$MeV (See also Fig.\ref{Bands2}(a)).
If the $J_z^2$-term is neglected (Fig.\ref{Kvalue}(b)), high-K states
with $K >10$ are not produced. 

Fig.\ref{Kvalue}(c) displays the results obtained 
without the residual two-body interaction (i.e.  only the
unperturbed cranked-Nilsson energy and the $J_z^2$-term are
considered).  Essentially, each solution obtained without the residual
interaction forms a rotational band, and each band has a value of
$K_i$ which changes only little as the spin changes.  We call these
solutions {\it the K-basis bands} (and denote by the letter
$\ket{\kappa}$ ) since they represent unperturbed rotational bands
where the K-quantum number is taken into account but the residual
interaction is neglected.  Over the whole energy region covered in
Fig.\ref{Kvalue}(c), 
the value of $K_i$  for the K-basis bands fluctuates from state
to state within a wide range, reaching values up to 
$K \sim 15$ (note that $K_i$ is positive  by definition).

With the residual interaction included (Fig.\ref{Kvalue}(a) 
vs. Fig.\ref{Kvalue}(c)), the number of 
high-K states with $K >10$ decreases in the high
excitation region $U>1.5$ MeV.  In fact, when the mixing 
becomes strong, the basis bands are mixed more and more democratically and 
the values of $K_i$ of individual
eigenstates tend to converge around an average value $K \sim 7$, 
with increasing $U$ as is seen in Fig.\ref{Kvalue}(a).

It is possible to describe the K-mixing more
quantitatively.  For this purpose we first consider a quantity which 
measures the degree of K-mixing within the individual 
energy levels.  Such a quantity may be defined in terms of the variance of the
operator $J_z^2$ for a given energy eigenstate $\ket{i}$, which is equal to 
$\bra{i}J_z^4\ket{i}- \bra{i}J_z^2\ket{i}^2$. We denote the average
value of this quantity by $\sigma^2(K^2)_{mix} =
\overline{\bra{i}J_z^4\ket{i}- \bra{i}J_z^2\ket{i}^2}$, where the bar
implies averaging over $i'$s in an energy bin.  

On the other hand, each energy level has a different  value of $K_i$
as can be seen  from the distribution of $K_i$ shown in Fig.\ref{Kvalue}(a).
We then consider the variance of this statistical fluctuation 
of $K_i$ among the different states $\ket{i}$, $\sigma^2(K^2)_{stat}$ =
$ \overline{\bra{i}J_z^2\ket{i}^2- \overline{\bra{i}J_z^2\ket{i}}^2}$.

Using these variances, a quantitative measure
of the degree of K-mixing can be given by the  ratio
\begin{equation} 
r_{Kmix} = { \sigma^2(K^2)_{mix} \over \sigma^2 (K^2)_{mix} +
\sigma^2(K^2)_{stat} }     \label{Kratio}
\end{equation}
between the variances of the intrinsic K-mixing and total
K-quantum number fluctuation. If the K-mixing is weak, the ratio 
$r_{Kmix}$ will attain a value close to zero.  
If many different K-quantum numbers are mixed strongly into individual 
levels in a random manner, $\sigma^2(K^2)_{mix}$ dominates and the 
ratio will take a value close to unity. As a guideline, 
one may define regions of weak K-mixing by $r_{Kmix} < 0.5$, and strong
K-mixing by $r_{Kmix} > 0.8$.
In accordance with the discussion of rotational damping in terms of
the average branching number, with the value $n_{branch} \approx 2$
defining the energy for onset of damping, an equivalent condition
for the onset of K-mixing should be $r_{Kmix} \approx 0.5$ (see Appendix
B).

Fig.\ref{Variance}(a) shows the two variances $\sigma^2(K^2)_{mix}$
and $\sigma^2(K^2)_{stat}$, and the total variance of the K-quantum
number as functions of $U$ for the states at spins 
$I=\frac{59}{2},\frac{61}{2}$.  
The K-mixing ratio calculated from these
variances is plotted in Fig.\ref{Variance}(b). 
The figure displays an onset of 
the K-mixing (as defined by $r_{Kmix}\approx 0.5$) around 
$U\approx 1.5$ MeV,  followed by
a monotonic approach towards complete K-mixing with increasing $U$.
It was noted above that the onset of rotational damping takes place around
$U \approx 1.0$MeV \cite{Herskind92,Dossing96}.
This onset energy
is predicted by calculations \cite{Lauritzen86,Matsuo97} and observed in
experiments in the rare earth nuclei around $^{168}$Yb. 
For $I=\frac{59}{2},\frac{61}{2}$, K-mixing sets in at higher energies
than rotational damping. Also, one
can observe in Fig.\ref{Kvalue}(a) that several high-K rotational 
bands remain discrete in the energy region $U=1.0-1.5$MeV, where 
most low-K states do not form rotational bands due to the rotational 
damping. Thus, it appears that the delayed onset of K-mixing, 
relative to the onset of damping, is also accompanied by a delayed
onset of damping of high-K bands.

\begin{figure}[t]
\centerline{\psfig{figure=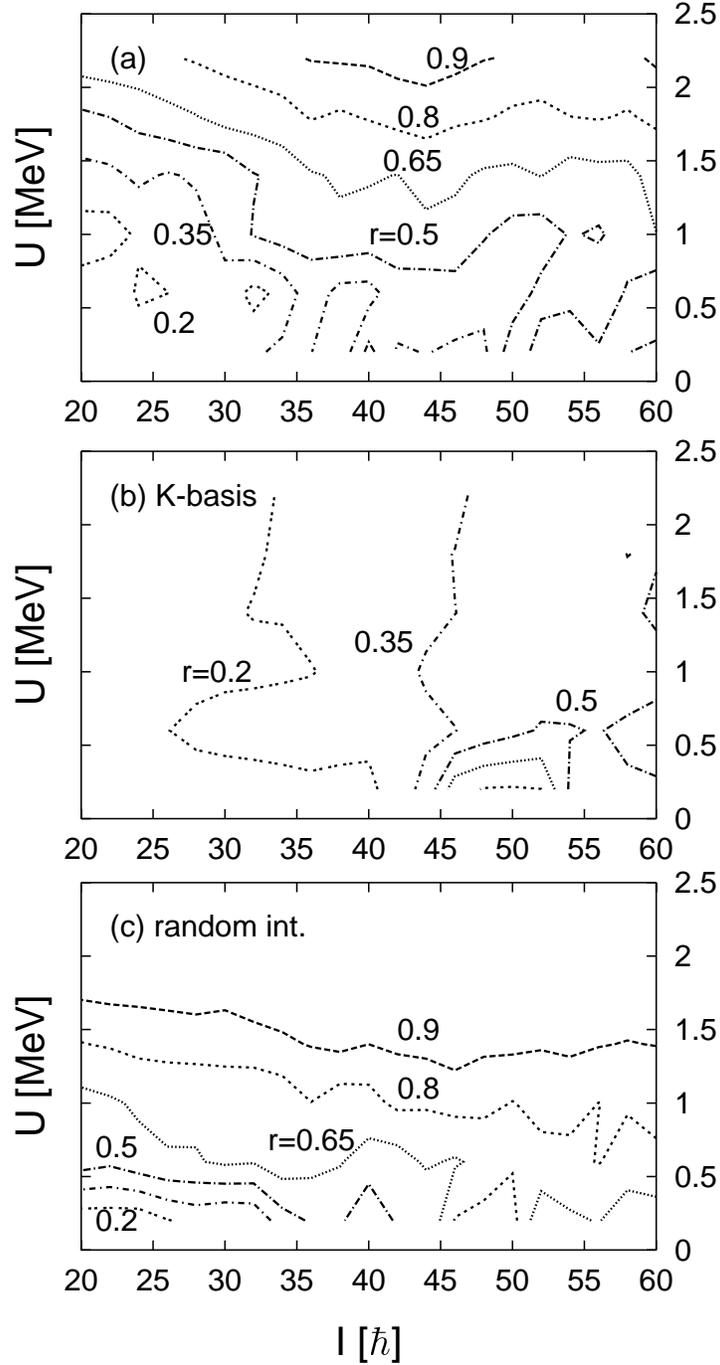,width=100mm}}
\caption{(a) Contour plot of the K-mixing ratio $r_{Kmix}$
as a function of spin $I$ and
thermal excitation energy $U$. (b) The same as (a), but the residual
SDI is neglected. (c) The same as (a), but the residual
SDI is replaced with a random two-body interaction.}
\label{Ratio}
\end{figure}

To obtain an overall view, the K-mixing ratio $r_{Kmix}$ is shown in 
Fig.\ref{Ratio}(a) as a function of spin and thermal excitation energy 
$U$.  It is seen that the K-mixing ratio increases 
from $r_{Kmix} \approx 0.2$ to $\approx 0.9$ as $U$
increases from zero to $U\approx 2$ MeV, and that the amount
of K-mixing increases with increasing angular momentum. 
The statistical limit ($r_{Kmix}\rightarrow 1$), is gradually approached
for higher thermal energies, $U > 2$ MeV. 

The delayed onset of K-mixing relative to damping displayed by figures
\ref{Kvalue}(a) and \ref{Variance}(a) is seen only at low
angular momenta $I \leq 35 \hbar$.  
The fact that K-mixing is rather weak even for rather high energy 
$U<1.5-1.8$ MeV at the lower
angular momenta may provide a first
explanation of the experimental observation \cite{Bosetti96} of the
persistence of the K-quantum number in the quasi-continuum spectrum
observed after a fusion reaction.

The role of the residual interaction in the K-mixing should be stressed.
In our calculations we have adopted the so-called surface delta interaction
(SDI), 
which respects the K-quantum number. Thus, the Coriolis term is essential
in order to induce K-mixing.
On the other hand, without the residual interaction,
the K-mixing  is very weak over the whole region considered, as can be seen 
in Fig.\ref{Ratio}(b), which corresponds to the  K-basis bands of 
Fig.\ref{Kvalue}(c).    
The Coriolis term alone only produces a weak K-mixing, gradually
increasing towards the very high spins $I=40-60$.
In other words,
the strong K-mixing ($U>2$MeV) in Fig.\ref{Ratio}(a) is a consequence of the
interplay between the two-body residual interaction, which is responsible for 
the order-to-chaos transition, 
and the Coriolis interaction which by itself would induce only 
a modest violation of the axial symmetry.

To further investigate the role of the two-body interaction, we show
in Fig.\ref{Ratio}(c) the result of a calculation where the SDI
is replaced by a random two-body interaction.
In this case, the two-body matrix elements are given by Gaussian random
numbers with the r.m.s. value of 12keV. This r.m.s. value is chosen to
reproduce the same onset energy of rotational damping as the SDI,
whose matrix elements have the r.m.s. value of 19 keV\cite{Matsuo97}.
The K-mixing and its energy dependence are much stronger 
than for the SDI interaction;
K-mixing sets in already at $U \sim 0.5$MeV, that is below 
the onset of rotational
damping, and strong K-mixing is already achieved at a much lower energy 
$U \sim 1.2$MeV.
The delayed onset of K-mixing seen at the low spins disappears with the
random interaction. 
These differences between the SDI and the
random interaction can be related to the fact that the SDI in itself
respects the K-quantum number, so the Coriolis interaction is needed
to induce K-mixing. Conversely, the K-quantum number is completely
ignored by the random interaction, and in this case, the degree of 
K-mixing is approximately given by the average 
number $N_{mix}$
of components of the wave function, as 
$r_{Kmix}\approx 1 - \frac{1}{N_{mix}}$ (cf. Appendix B).

\section{K-selection in interaction matrix elements}

For resolved bands of low thermal excitation energy, some interaction matrix 
elements may be determined at band crossings of pairs of bands having 
different rotational frequency. Recently, such experimental interaction matrix 
elements were extracted from an analysis of band crossings in 
$^{163}$Er\cite{er163} and its neighbors. In order to investigate
K-selection rules associated to these matrix elements, 
one may assign the K-value
from the band head spin, and plot the extracted interaction strength 
versus the difference $\Delta K$ in K-quantum numbers for the crossing bands. 
(One can justify this assignment by relating to our cranked shell
model, where we find that the K-values calculated in the cranked shell 
model are not identical to the band head K-values, but they are 
on the other hand not very different for the unmixed rotational bands 
at the lowest heat energies.)

\begin{figure}[t]
\centerline{\psfig{figure=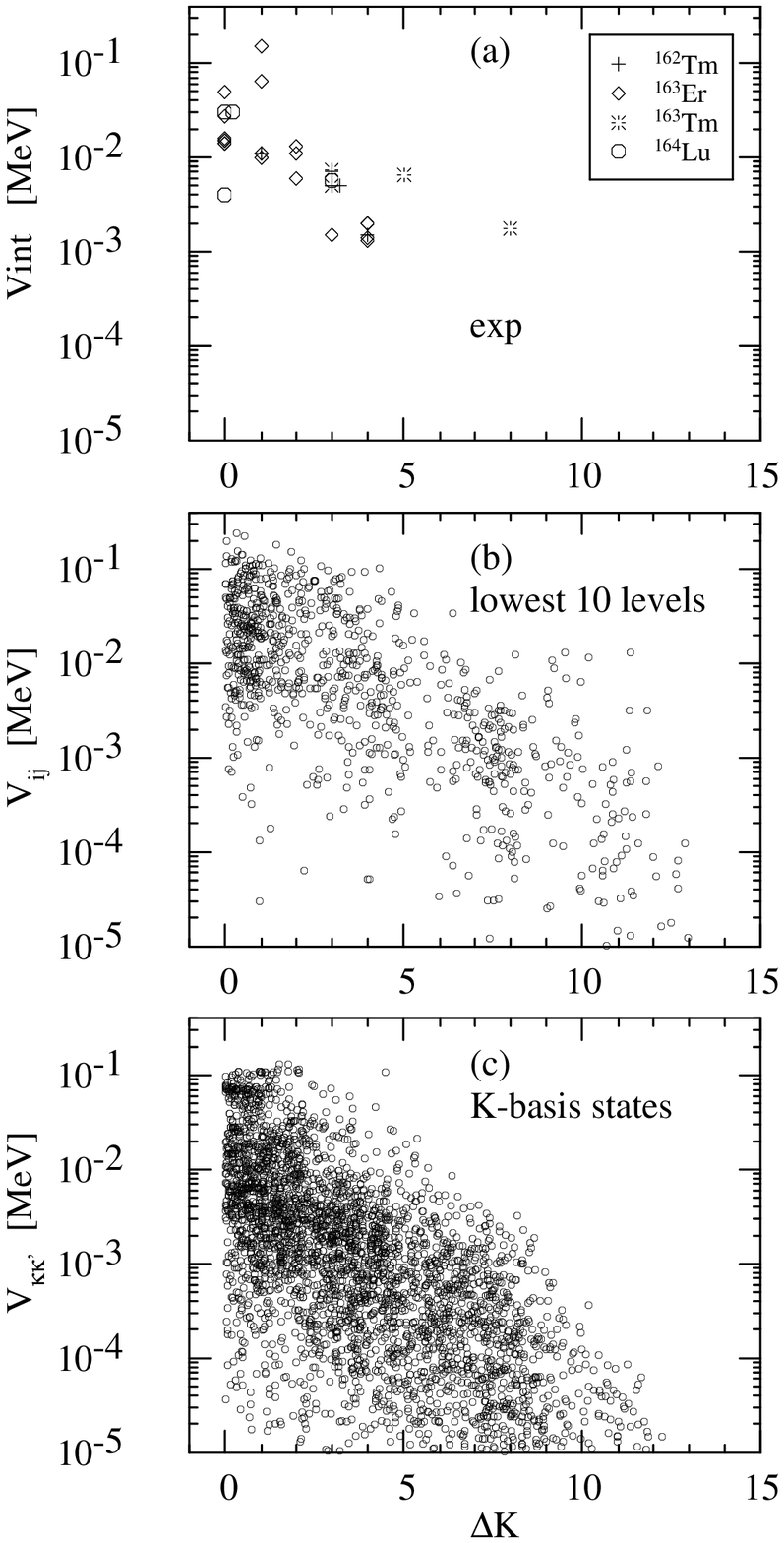,width=90mm}}
\caption{
(a) The experimentally
extracted interaction strengths at band crossings, correlated with
the difference $\Delta K$ between the band-head K-quantum numbers
of the crossing bands (see text).
(b) The calculated matrix elements $V_{ij}$ of the residual
SDI among
the lowest energy levels(10 for each $I^\pi$), which approximately
correspond to the states forming discrete rotational bands, 
where $V$ is its absolute size while $\Delta K$ is the difference
of K values of two energy levels. Those calculated for
$I^\pi={39/2}^{\pm} - {57/2}^{\pm}$ are collected. 
(c) The calculated matrix elements $V_{\kappa\kappa'}$ 
among the lowest 20 K-basis states.
}
\label{Matel}
\end{figure}

The result obtained from the 32 band crossings 
in $^{163}$Er\cite{er163} and neighboring nuclei $^{162}$Tm \cite{tm162},  
$^{163}$Tm\cite{tm163}, and $^{164}$Lu \cite{lu164}
is shown in Fig.\ref{Matel}(a). 
Nineteen crossings can be related to 
a two-body residual interaction, and a previous analysis 
limited to them \cite{Dossing99} already found that they correspond to
an average interaction strength of 14 keV, and that they are
correlated with $\Delta K$. The other crossings include
those caused by the changes in the pair-field with increasing 
rotational frequency, which are not well described in the present model.
The distribution of the the experimental matrix elements
suggests an exponential scaling 
\begin{equation}\label{Int-K}
V_{K K'} \propto
\exp(-|K-K'|/\delta K_0)
\end{equation}
with respect to the difference in K between the crossing bands, and
from figure \ref{Matel}(a)  one may roughly read off an effective 
K-correlation interval $\delta K_0 \approx 1.6$.

We compare the experimental matrix elements to matrix elements of the 
SDI among the energy eigenstates in the region of resolved bands. 
Fig.\ref{Matel}(b) plots the matrix elements $V_{ij}$
calculated for all pairs of the lowest 10 energy levels for each $I^\pi$,
approximately corresponding to an interval in thermal energy $U=0-1$ MeV. 
Comparing panels (a) and (b) of the figure, one sees that the 
calculations are in accordance with 
the main features of the experiment: $(i):$ 
a median around $\sim 20$keV at $\Delta K = 0$, $(ii)$ a considerable 
fluctuation among matrix elements at the same $\Delta K$, and 
$(iii)$ an overall exponential decrease with increasing $\Delta K$. 
For the calculated low lying energy
eigenstates, we find a somewhat weaker K-selection rule than that
displayed by the experimental band crossings, corresponding
to a wider K-correlation interval $\delta K_0 \approx 2.1$.

The experimental matrix elements are extracted from energy level
repulsion and E2 branching ratio between two crossing bands while
the theoretical values plotted in Fig.\ref{Matel}(b) are the matrix
elements calculated directly from the energy eigenstates, which
sometimes become a mixture of rotational bands. The small difference in 
the K-correlation extracted from  Figs.\ref{Matel}(a) and (b) could in part 
originate from the different methods adopted to extract the matrix elements.

As another possible evaluation of matrix elements,
one can also consider the
matrix elements of the SDI residual interaction among the lowest 
energy K-basis states. This is done in fig.\ref{Matel}(c).
The calculated matrix elements with the K-basis 
display the same dispersion for fixed $\Delta K$, and 
a more steep slope with $\Delta K$, corresponding to 
a correlation interval $\delta K_0 \approx 1.5$. 
Nevertheless, both the K-correlation intervals 
calculated by the two different definitions agrees fairly well
with the experimental value.

Since it is the Coriolis force that eventually breaks the
K-quantum number, the above K-selectivity of the matrix elements
becomes weaker as the spin increases.  For very high spins
$I=50-58$, the K-correlation interval is found to be $\delta
K_0 \sim 3-4$. Altogether, 
one expects a linear correspondence between $\delta K_0$ and the angular 
momentum for the low-lying bands. Going to the opposite
extreme, the low angular momenta close to the band heads, the
typical K-forbiddeness factor \cite{BM2} 
of $10^{-2}$ per unit change of $K$ 
of transition strengths translates into a K-forbiddeness factor of $10^{-1}$ 
for matrix elements, and thereby $\delta K_0 \sim 0.4$. 

\section{Effective number of bands}

The present model calculations should also address the results found in
quasicontinuum spectra gated on specific
low-lying bands having a well-defined  K-quantum number\cite{Bosetti96}.
However, a quantitative comparison between theory and experiments 
requires a detailed description of the $\gamma-$cascades. This will be 
attempted in a future publication, in which a comparison can be made
to new quasicontinuum data. The mixing coefficient $r_{Kmix}$ will
be an important ingredient also for understanding the
K-selection rules displayed by the cascades. 

Here, we restrict ourselves to the number of paths $N_{path}$ 
associated to low-K or high-K bands. 
These quantities can be extracted by a fluctuation analysis from the 
ridge part of the two-dimensional $E_{\gamma 1} \times E_{\gamma 2}$ spectra.
$N_{path}$ represents an effective number of regular rotational bands 
which are populated by the gamma-cascades. 

As indicated in Fig.\ref{Bands}, the number of discrete regular
rotational bands present in the nucleus will be rather small because of 
rotational damping. A regular rotational band which gives rise to counts
on the ridge in two-dimensional spectra should contain two strong consecutive
transitions, obeying rotational energy correlations, that is they
should be separated by about $4/{\cal I}$, ${\cal I}$ being a typical
band moment of inertia. In our calculations, 
the number of bands $N_{band}$ at a given spin 
is defined as the total number of levels
for which  $n_{branch}<2$ \cite{Matsuo97}. Here, 
$n_{branch}=(\sum_j s_{ij}^2)^{-1}$ is an effective number of E2 
decay-branches where $s_{ij}$ is denotes  the normalized E2 
transition strength for going from the level $i$ at spin $I$ to the level
$j$ at spin $I-2$.

\begin{figure}[t]
\centerline{\psfig{figure=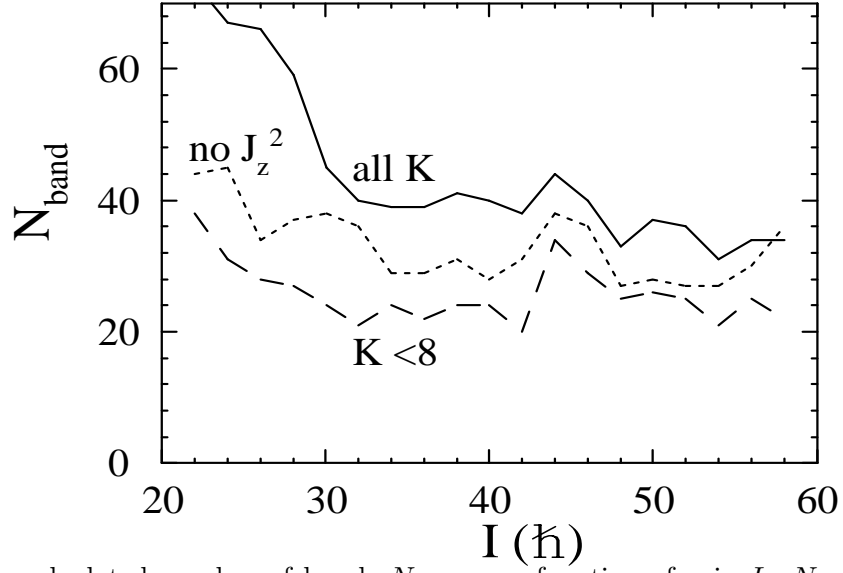,width=110mm,angle=-90}}
\caption{The calculated number of bands $N_{band}$ as a function of spin $I$.
$N_{band}$ for the states with $K<8$ is plotted by the
dashed curve. $N_{band}$  calculated without the $J_z^2$-term is also
shown with dotted curve.}
\label{Nband}
\end{figure}

The calculated values of $N_{band}$ summing over four configurations 
of parity and signature are shown in Fig.\ref{Nband}. In order to 
have some insight in the role played by
the K-quantum number, we also separately evaluate $N_{band}$ associated with
high-K and low-K states, which are classified with the values of 
$K_i=\sqrt{\bra{i}J_z^2\ket{i}}$. The total $N_{band}$ is around 
$\approx 30-40$ except for $I<30$. This is consistent with the
experimental value of $N_{path}\approx 40$\cite{Bosetti96} obtained
for ungated spectra. 

Rare-earth nuclei previously analysed with
respect to ridge fluctuations, $^{168}$Yb, $^{163}$Tm, and 
$^{164}$Yb, were found to possess around $\sim 25$ 
regular rotational bands\cite{Herskind92,Dossing96}. 
It appears that the surplus of bands in $^{163}$Er are actually 
high-K bands, which do not exist in the other nuclei. 
An experimental 
value of $N_{path}\approx 15$ is obtained for the spectrum gated on a 
high-K band with $K=19/2$ \cite{Bosetti96}.
Fig.\ref{Nband} shows that the number of low-K bands $(K<8)$ is around 
25, and the number of high-K bands $(K>8)$ is around 15.
Here the value $K=8$ distinguishing between
low-K and high-K bands is somewhat arbitrarily taken as the upper limit
of K for the bands calculated without the $J_z^2$-term, cfr. fig
\ref{Kvalue}(b).  Fig.\ref{Nband} also shows the number of bands
calculated without the $J_z^2$-term. It is seen that 
high-K rotational bands generated by the $J_z^2$-term 
increase the total number of bands by about $\sim 10$. 

\section{Conclusions}

We have discussed the onset of K-mixing 
in rapidly rotating warm  nuclei by means of microscopic
cranked shell model calculation performed for the nucleus $^{163}$Er.  

Although the K-mixing due to the Coriolis force alone is generally weak,
it is enhanced by the configuration mixing produced by 
the residual interaction among the cranked mean-field np-nh bands,
becoming more pronounced as the level density increases.
A quantitative measure of K-mixing can be given by the K-mixing ratio
introduced above. 
For angular momenta up to about 35$\hbar$, the onset of K-mixing 
takes place around a heat energy of $U \approx 1.5-1.8$ MeV, while
the statistical limit of strong K-mixing is approached only 
above $U>2$ MeV. 
This  dependence of K-mixing on excitation energy 
is related to the K-selectivity of the
matrix elements of the two-body residual interaction we have adopted, 
which seems to be consistent with the  matrix elements 
experimentally observed at the band crossing interactions. 
The onset of K-mixing is retarded compared to the onset of damping 
which, for the same surface-delta interaction, occurs around $U \approx
1$ MeV. At higher angular momenta $I \ge 35 \hbar$, 
K-mixing and damping set in around the same heat energy of
$U \approx $ 1 MeV.
In the case of a purely random
interaction, the K-mixing is much stronger, and 
essentially complete already at $U \approx $ 1 MeV.

Theoretical results from the same type of calculations may in
the future be used for  an analysis of the gamma-ray cascades,
thereby addressing the persistence of the K-quantum  number, 
found in the experimental  analysis of
quasi-continuum spectra following fusion reactions \cite{Bosetti96}.

\section*{Appendix A. High-K states in the signature basis}

Our shell model basis $\ket{\mu(I)}$ consists of 
the single-particle orbits obtained diagonalizing
the cranked single-particle Hamiltonian
$h_{cr}=h_{Nilsson}-\omega_x J_x$. The Hamiltonian
is symmetric with respect to a rotation of angle $\pi$ 
about the $x$-axis, so that the single-particle orbits have 
the signature as a good quantum
number. The shell model basis $\ket{\mu(I)}$ also keeps the signature,
but on the other hand, does not have a good K-quantum number,
especially for many-particle many-hole configurations.
Diagonalization of the $J_z^2$-term added to $H$
in the cranked shell model Hamiltonian, Eq.(\ref{smHamlK}), is 
crucial to produce the high-K states.

To illustrate, 
let us consider a subspace of
one-particle one-hole configurations made of two deformation-aligned
orbits, which have single-particle K-quantum numbers $\Omega_p>0$ and
$\Omega_h>0$.  
To each orbit are associated two almost degenerate states 
$\phi_{i\pm}={1\over \sqrt{2}}(\phi_{i, K=\Omega_i}
\pm\phi_{i,K=-\Omega_i})$
with both signatures. (Here $\phi_{i, K=\Omega_i}$ and
$\phi_{i, K=-\Omega_i}$ carry positive and negative value for the
K-quantum number operator $J_z$.) Correspondingly,
the 1p1h configuration has four degenerate cranked shell-model basis states 
$\phi_{p+}\phi_{h+}^{-1}, \phi_{p-}\phi_{h-}^{-1}={1\over 2}\left(
\phi_{p,K=\Omega_p}\phi_{h,K=\Omega_h}^{-1}+
\phi_{p,K=-\Omega_p}\phi_{h,K=-\Omega_h}^{-1}\right)
\pm{1\over 2}\left(
\phi_{p,K=\Omega_p}\phi_{h,K=-\Omega_h}^{-1}+
\phi_{p,K=-\Omega_p}\phi_{h,K=\Omega_h}^{-1}\right)
$ and
$\phi_{p+}\phi_{h-}^{-1}, \phi_{p-}\phi_{h+}^{-1}={1\over 2}\left(
\phi_{p,K=\Omega_p}\phi_{h,K=\Omega_h}^{-1}-
\phi_{p,K=-\Omega_p}\phi_{h,K=-\Omega_h}^{-1}\right)
\pm
{1\over 2}\left(\phi_{p,K=\Omega_p}\phi_{h,K=-\Omega_h}^{-1}-
\phi_{p,K=-\Omega_p}\phi_{h,K=\Omega_h}^{-1}\right).$ 
Note that none of these 1p1h basis states
has pure $K=\Omega_p + \Omega_h$ or 
$K=|\Omega_p - \Omega_h|$.
Once the $J_z^2$-term is
diagonalized in the subspace, these states are re-coupled
to produce eigenstates of $J_z^2$. These eigenstates receive energy
contributions $-J_z^2/2\calJ \sim -(\Omega_p\pm \Omega_h)^2/2\calJ$, yielding
two energetically favored high-K states 
${1\over\sqrt{2}}
(\phi_{p,K=\Omega_p}\phi_{h,K=-\Omega_h}^{-1}\pm
\phi_{p,K=-\Omega_p}\phi_{h,K=\Omega_h}^{-1})$ 
 with $K=\Omega_p + \Omega_h$ and different signature, 
and two unfavored low-K
states 
${1\over\sqrt{2}}
(\phi_{p,K=\Omega_p}\phi_{h,K=\Omega_h}^{-1}\pm
\phi_{p,K=-\Omega_p}\phi_{h,K=-\Omega_h}^{-1})$ 
with $K=|\Omega_p - \Omega_h|$.  The same mechanism applies to
many-particle and many-hole configurations, for which the eigenstate
with the highest K configuration $K=\Omega_{p1} + \Omega_{p2}+ \cdots +
\Omega_{h1}+\Omega_{h2}+\cdots$ is produced as the most energetically favored
state. We have found that  most high-K states obtained in the numerical
calculation have a similar structure, 
although the K-quantum number is only approximately conserved due to  
the Coriolis effect (i.e. to  the cranking
term).

\section*{Appendix B. Random K-mixing}

We consider here the K-mixing ratio $r_{Kmix}$ in the statistical
limit, for which the expansion coefficients of wave functions
on K-basis states will be statistically independent.
An energy eigenstate is expanded on the K-basis
\begin{equation}
\ket{i} = \sum_n a_n \ket{\kappa_n} 
\end{equation}
with amplitudes $a_n$ and the corresponding probabilities
$P_n = |a_n|^2$. $J_z^2$ is diagonal in that basis:
\begin{equation}
\bra{i} J_z^2 \ket{i} = \sum_n |a_n|^2 K_n^2 = 
\sum_n P_n K_n^2, \hspace{1cm} 
\bra{i} J_z^4 \ket{i} = \sum_n P_n K_n^4. 
\end{equation}
Averaging over the mixed band states, one obtains
\begin{equation}
\overline{\bra{i} J_z^4 \ket{i}} = \overline{\sum_n P_n K_n^4} 
= \overline{K_b^4} 
\end{equation}
where the last term denotes averaging over the basis bands.
Further, one obtains:
$$
\overline{\bra{i} J_z^2 \ket{i}^2} =
\overline{(\sum_n P_n K_n^2)^2} = 
\overline{\sum_n P_n^2 K_n^4 + \sum_{n \neq n'} P_n P_n' K_n^2 K_n'^2} \\
=\overline{\sum_n P_n^2 K_n^4} 
+ \left(\overline{\sum_n P_n K_n^2}\right)^2
- \sum_n \left(\overline{ P_n K_n^2}\right)^2.
$$
Here, the statistical independence between the different components
of basis bands has been applied.

Defining the average mixing number into the states
$$
\frac{1}{N_{mix}} \equiv \overline{\sum_n{P_n^2}},
$$
one obtains the variance of $J_z^2$ in mixed states:
$$
\sigma^2(K^2)_{mix}=
\overline{\bra{i} J_z^4 \ket{i}} - \overline{\bra{i} J_z^2 \ket{i}^2} =
\left(1-\frac{1}{N_{mix}}\right)
\left[\overline{K_b^4} - \overline{K_b^2}^2 \right].
$$
The variance of statistical fluctuation of $K^2=\bra{i} J_z^2 \ket{i}$
is obtained by inserting the same
results:
\begin{equation}
\sigma^2(K^2)_{stat}=
\overline{\bra{i} J_z^2 \ket{i}^2} - \overline{\bra{i} J_z^2 \ket{i}}^2 
= \frac{1}{N_{mix}} \left[\overline{K_b^4} - \overline{K_b^2}^2 \right].
\end{equation}
Thus, one finds that the K-mixing ratio, Eq.(\ref{Kratio}), is given by
\begin{equation}
r_{Kmix} = 1-\frac{1}{N_{mix}}.
\end{equation}
In fact, the Coriolis interaction obeys strong selection rules
on the matrix elements $\Delta K = \pm 1$, and with a residual
interaction, which obeys the K-quantum number, one will not expect
these statistical considerations to be valid. Rather, they are
relevant for a random residual interaction, and there one
sees that mixing even a small number 
of states causes strong K-mixing. Already a two-state mixing
$N_{mix}=2$ corresponds to $r_{Kmix} = 0.5$ and $N_{mix}=5$ is enough to 
generate strong K-mixing, $r_{Kmix} = 0.8$.\\

\vfill
\break

\vfill
\break

\vfill
\break

\vfill
\break
\begin{figure}[t]
\centerline{\psfig{figure=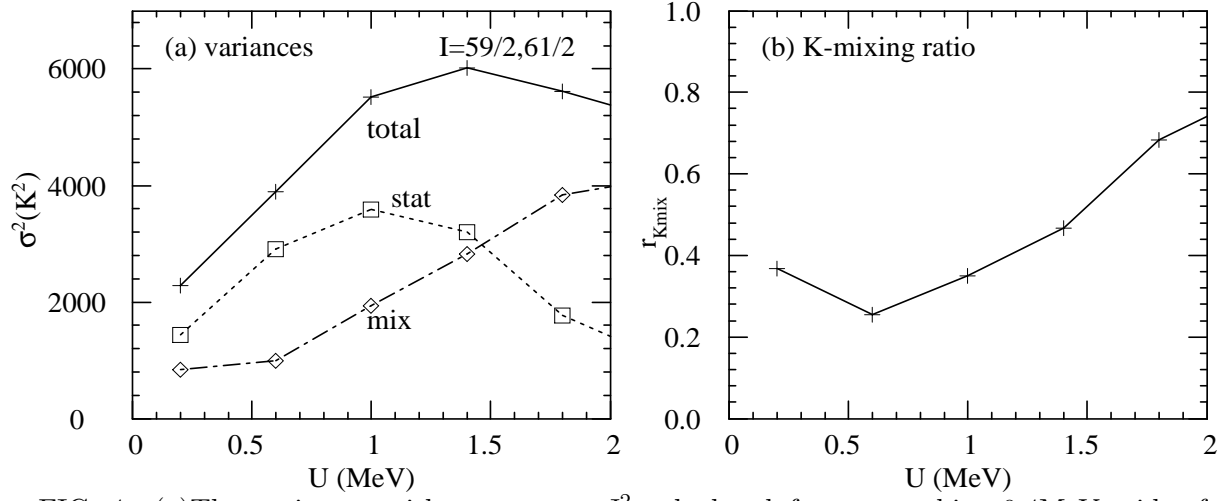,width=160mm}}
\caption{(a)The variances with respect to $J_z^2$ calculated for
energy bins 0.4MeV wide, for
spins $I=59/2,61/2$ including both parities.
The solid, dotted and dashed lines indicate 
$\sigma^2(K^2)_{tot}, \sigma^2(K^2)_{stat}$, and $\sigma^2(K^2)_{mix}$,
respectively. In this calculation, 10000 basis states are used
for diagonalization. (b) The K-mixing ratio $r_{Kmix}$ calculated 
for the same energy bins.}
\label{Variance}
\end{figure}

\vfill
\break

\vfill
\break

\vfill
\break

\end{document}